\documentclass[12pt,preprint]{aastex}
\usepackage{graphics}

\def\gtorder{\mathrel{\raise.3ex\hbox{$>$}\mkern-14mu
             \lower0.6ex\hbox{$\sim$}}}
\def\ltorder{\mathrel{\raise.3ex\hbox{$<$}\mkern-14mu
             \lower0.6ex\hbox{$\sim$}}}

\def\Msun{\>{\rm M_{\odot}}}

\shorttitle{Limits on Substellar Objects Around G29-38}
\shortauthors{Debes et al.}

\begin{document}
\title{Cool Customers in the Stellar Graveyard I: Limits to Extrasolar Planets around the White Dwarf G29-38}
\author{John H. Debes\altaffilmark{1}, Steinn Sigurdsson\altaffilmark{1},
Bruce E. Woodgate\altaffilmark{2}}

\altaffiltext{1}{Department of Astronomy \& Astrophysics, Pennsylvania State
University, University Park, PA 16802}
\altaffiltext{2}{NASA Goddard Space Flight Center, Greenbelt, MD 27710}

\begin{abstract}
We present high contrast images of the hydrogen white dwarf
G 29-38 taken in the 
near infrared with the Hubble Space Telescope
and the Gemini North Telescope as part of a high contrast imaging search for
substellar objects in orbit around nearby white dwarfs.
  We review the current limits on planetary companions
for G29-38, the only nearby white dwarf with
an infrared excess due to a dust disk.   We add our recent observations
 to these
limits to produce extremely tight constraints
 on the types of possible companions 
that could be present.  No objects $>$ 6 M$_{Jup}$ are
 detected in our data
at projected
separations $>$ 12 AU, and no objects $>$ 16 M$_{Jup}$ are detected for separations from
3 to 12 AU, assuming a total system age of 1 Gyr.  Limits for companions
at separations $<$ 3 AU come from a combination of 2MASS photometry and
previous studies of G29-38's pulsations.   Our imaging with Gemini
 cannot confirm a 
tentative claim for the presence of a low mass brown dwarf.   These observations demonstrate
that a careful combination of several techniques can probe
 nearby white dwarfs for large planets and low mass brown dwarfs.

\end{abstract}   

\keywords{circumstellar matter --- planetary systems --- white dwarfs --- stars: individual (G29-38)}

\section{Introduction}
\label{s1}
G29-38 (ZZ Psc, WD 2326+049, GJ 895.2) is a nearby ($d$=13.6 pc) non-radially pulsating hydrogen 
white dwarf (WD) with photospheric absorption lines due to metals such as Mg and
Ca \citep{vanaltena95,koester97}.  Hydrogen WDs with metal absorption
lines are known as DAZs. 
G29-38 has a measured gravity $\log{g}$= 8.15 and a T$_{eff}$=11820 K,
 placing its cooling age at 0.6 Gyr \citep{liebert04}.  

G29-38 possesses an infrared excess, originally 
attributed to a companion substellar object \citep{zuckerman87}.
  Further infrared studies,
including pulsational studies in the near-IR, showed that the excess
was more consistent with a
circumstellar disk at 1 R$_\odot$ with a blackbody temperature of $\sim$1000~K
\citep{tokunaga88,tokunaga90,telesco90,graham90}.  The origin of the disk is unclear, though it could be caused by a tidally
 disrupted asteroid or comet, potentially sent to the inner system
by a
planetary system that suffered chaotic evolution after post main sequence evolution \citep{debes02,jura03}.  

Long-term pulsational studies of G29-38 have allowed 
several of the more stable pulsation modes to be monitored for timing delays due to
an unseen companion \citep{kleinman94,kleinman98}.  No conclusive
 detection of a companion has been
reported.  Speckle imaging of 
G29-38 furthermore could not detect any unresolved companions, although IR
slit scans of G29-38 appeared to show an extension in the N-S direction on 
scales of 0.4\arcsec\ \citep{kuchner98,haas90}.

The biggest question that remains is the origin of the dust disk, which
pollutes the white dwarf's atmosphere with metals.  Any origin for the dust 
requires a substellar companion \citep{debes02,zuckerman03}.
 Planets
in inner regions most likely are engulfed by the AGB phase of the
star, with larger planets possibly ``recycled'' into brown dwarf companions
\citep{seiss99a,seiss99b}.  Remnant
asteroids and comets potentially could survive at 
distances where they would not be ablated during the AGB phase \citep{stern90}.
  However,
if the primary star has asymmetric mass loss, objects such as comets
can easily be 
lost from the system if the orbital timescale equals the 
timescale for mass loss \citep{parriott98}. 
Planets or brown dwarfs in orbits $\gtorder$5~AU will avoid engulfment 
and survive post
main sequence evolution \citep{rasio96,duncan98}.  
  Massive white dwarfs that are the result of WD-WD mergers
may also form terrestrial mass 
planets in the debris of the merger, allowing unseen companions in close 
orbits \citep{livio92}.

WDs also make excellent targets for extrasolar planet searches with
current ground and space based techniques \citep{burleigh02,debes04}.
WDs are orders of magnitude dimmer than their main sequence progenitors, 
allowing fainter companions to be detected.  In the near-IR substellar
 companions
emit thermal radiation, which for objects warmer than $\sim$300 K
dominates the reflected
 light from their hosts.
Companions that form at a particular semi-major axis conserve angular momentum
 during post main sequence mass loss and widen their orbits by a factor
 $\propto m_i/m_f$, where $m_i$ and $m_f$ are the initial and final masses of the 
central star \citep{jeans24}.  Any observations of a WD then probe to
orbits that were a factor of at least 2 times smaller when the star was on the
main sequence.  
Current imaging searches in the near infrared
 are most effective for WDs that have a combined cooling time and main sequence
 age of
$\sim$1-5 Gyr.  At these ages WDs have become dimmer than their main sequence
progenitor.  Concurrently, massive planets and brown dwarfs are 
observable in the near-IR since they haven't cooled below 300 K.
  WDs with metal lines can be markers for planetary systems and the
presence of a dust disk and a high abundance of accreted metals
makes G29-38 a primary candidate for the presence
of a substellar or
planetary companion \citep{debes02}.   

These motivations are the basis for a survey of nearby young DAZs that
 we have conducted using the Hubble Space Telescope (HST).  We have primarily
used the coronagraph on the NIC2 detector which is part of NICMOS.
  With the high contrast, resolution, and sensitivity of
NICMOS, we can probe to within 3 AU of G29-38 looking for substellar
companions that could help to explain the presence of this peculiar
DAZ's dust disk.  Section \ref{s2} describes the observations. 
Section \ref{s3} presents sensitivity limits as well as second
epoch data for a candidate companion.
  These results are then combined
with pulsational timing studies and 2MASS photometry 
 to perform the most comprehensive search for
substellar companions around a WD to date, providing
a roadmap for the direct detection of planetary companions to WDs
in the future.  
In Section \ref{s4} we present the conclusions from our work.

\section{Observations}
\label{s2}

 We  imaged G29-38 using the NIC-2 camera on NICMOS both with and
without a coronagraph.  
We used both the F110W ($\sim$J) and F160W ($\sim$H) filters
for our observations.
The highest degree of contrast at separations $>$ 1\arcsec\ 
is gained by performing a combination
of coronagraphy and point spread function (PSF) subtraction \citep{fraquelli04}.  Pipeline reduced coronagraphic data were obtained from STScI, and the basic
 procedure outlined
by \citet{fraquelli04} was used to optimize the results for coronagraphic
self-subtraction.  

Due to the detection of a candidate planetary companion, follow-up observations
were taken approximately a year later with Gemini North telescope Director's 
Discretionary time.  We used the Altair adaptive optics (AO) system in conjunction
with NIRI to take H band images of G 29-38 and the candidate to determine if 
they shared common proper motion.  

The Gemini observations were taken on August 5, 2004.
A total of 4 $\times$ 15s frames were co-added at 10 dither points to
subtract the background and to remove pixel-to-pixel defects, for an effective 
integration on source of forty minutes.  Our total integration returned an
average AO corrected FWHM of 75 mas, significantly smaller than the diffraction limit of 
our F110W images with HST.  Because of Gemini's higher spatial resolution,
we used this second epoch data to search for companions at separations 
$<$1\arcsec.  Table \ref{tab:obs} shows the date and time of the observations
 taken of 
G29-38, along with the filters. 

The second epoch Gemini data were processed using several IRAF tasks designed by
the Gemini Observatory and based upon the samples given to observers.  Each
frame was flatfielded and sky subtracted.  In addition, due to the on-sky 
rotation from a fixed Cassegrain
 Rotator, each frame was rotationally registered
and combined.  More details of the general strategy and reduction are in
\citet{debes04}.

\section{Results}
\label{s3}

\begin{figure}
\plottwo{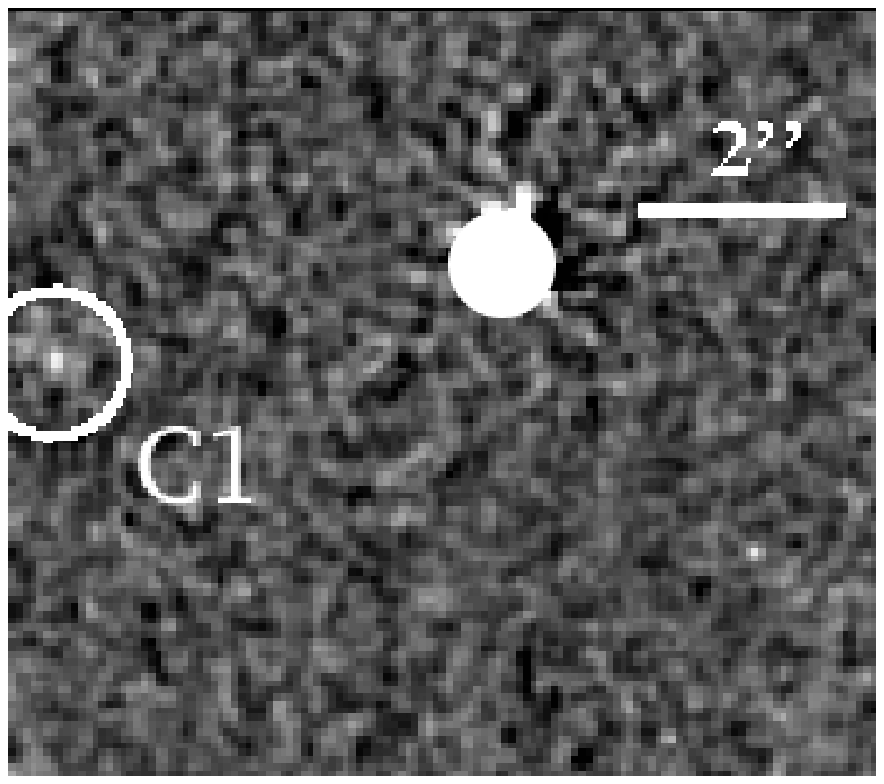}{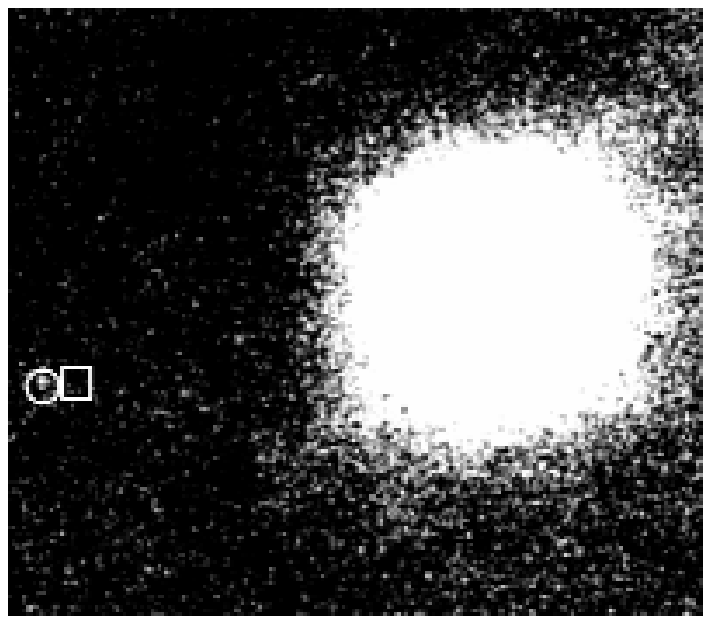}
\caption{\label{fig:gemfig} (left) Discovery image of a candidate planetary
 companion in the HST F160W filter.  The image was smoothed with a
Gaussian filter, C1 marks the candidate, and G29-38 is masked out.
 Other features are either subtraction artifacts or detector artifacts. 
(right) Second epoch image with Gemini,
along with the predicted positions of co-moving (square) and non co-moving
(circle) objects.  The object is non co-moving and therefore in the background. In both images North is rotated 36$^\circ$ clockwise.}
\end{figure}

No substellar objects were detected in an annulus betweee 1\arcsec\ and 
5\arcsec\ from G 29-38 with our coronagraphic observations.  
One candidate object was detected at a S/N$\sim$6 with m$_{F110W}$-m$_{F160W}$=1.1$\pm$0.3 and apparent m$_{F110W}$=23.7$\pm$0.2.  The discovery image and its follow up Gemini image is 
shown in Figure \ref{fig:gemfig}.  The magnitudes and colors were consistent with an object $<$ 10 M$_{Jup}$ at
13.6~pc \citep{bsl03}.  
Its initial position relative to G 29-38 was $\Delta\alpha$=4.91\arcsec$\pm0.01$
$\Delta\delta$=2.03\arcsec$\pm0.01$ in our HST images.
Since the measured proper motion of G29-38 is -411$\pm$0.01 mas/yr in $\alpha$ and -263$\pm$.01
mas/yr in $\delta$ \citep{pauli03}, we predict an increase of 330 mas and 250 mas in R.A. and dclination, respectively, between our two epoch observations due to parallactic motion and proper motion, leading to $\Delta\alpha$=5.24\arcsec$\pm$0.02 and $\Delta\delta$=2.28$\pm$0.02
 for the non
co-moving case.  The position of the candidate in the second epoch Gemini data
is  $\Delta\alpha$=5.25\arcsec$\pm$0.01 and $\Delta\delta$=2.30\arcsec$\pm$0.01.
 
 The candidate is a background object
that does not share G29-38's proper motion.  The errors in the calculation come
primarily from the uncertainty in G29-38's proper motion and uncertainties
in the measured centroids.  However, the position of the background object is 
well within the errors and shows no hint of its own proper motion.

 Our Gemini
data were of high enough spatial resolution that we should have easily detected
extended structure similar to what was reported in \citet{haas90}.  We see no such
structure in any of our HST or Gemini observations.  Any dust disk present
around G29-38 must be confined to smaller than 75 mas or 1 AU projected 
separation.

\subsection{Limits from Imaging}

\citet{schneider03} showed
 a reliable way to determine sensitivity of an observation with
NICMOS, given the stability of the instrument.  Artificial ``companions'' are
generated with the HST PSF simulation software TINYTIM \footnote{http://www.stsci.edu/software/tinytim/tinytim.html} and scaled to higher fluxes
until they are recovered.
  These companions are 
 inserted into the observations and used to gauge sensitivity.  We adopted
this strategy for our data as well.  An implant was placed in the images.
  Two difference
images were created following our procedure of PSF subtraction
and then rotated and combined for maximum signal to noise.
Sample images were examined
 by eye as a second check that the dimmest implants could be recovered.
The implants were normalized so that their total flux was equal to
1 DN/s.   The normalized value was converted to a flux in Jy or a Vega magnitude by 
multiplying by the correct photometry constants given by the NICMOS Data Handbook.   
We considered an implant recovered if its scaled
flux in a given aperture had a S/N of 5.   

 For our Gemini data we
used the PSF of G29-38 as a reference for the implant.  The implant was normalized to a peak pixel value of one.  Implants were scaled with increasing flux
until recovered to determine the final image's 
sensitivity to objects at a S/N of 10, since siginificant
 flux from the PSF remained at separations $<$ 1\arcsec.  The relative flux of the implant with respect
to the host star was measured and a corresponding MKO
 H magnitude was derived from the 2MASS H magnitude to give a final apparent
magnitude sensitivity.  For our Gemini images we checked sensitivity starting 
at a distance of $\sim$3 times the FWHM of G29-38, or  0.22\arcsec,
 out to 7\arcsec, the extent of our field of view.  Gemini's sensitivity beyond $\sim$1.5\arcsec\ was 
comparable to that of our NICMOS data, with a median sensitivity of H$\sim$22.9.

\begin{figure}
\plotone{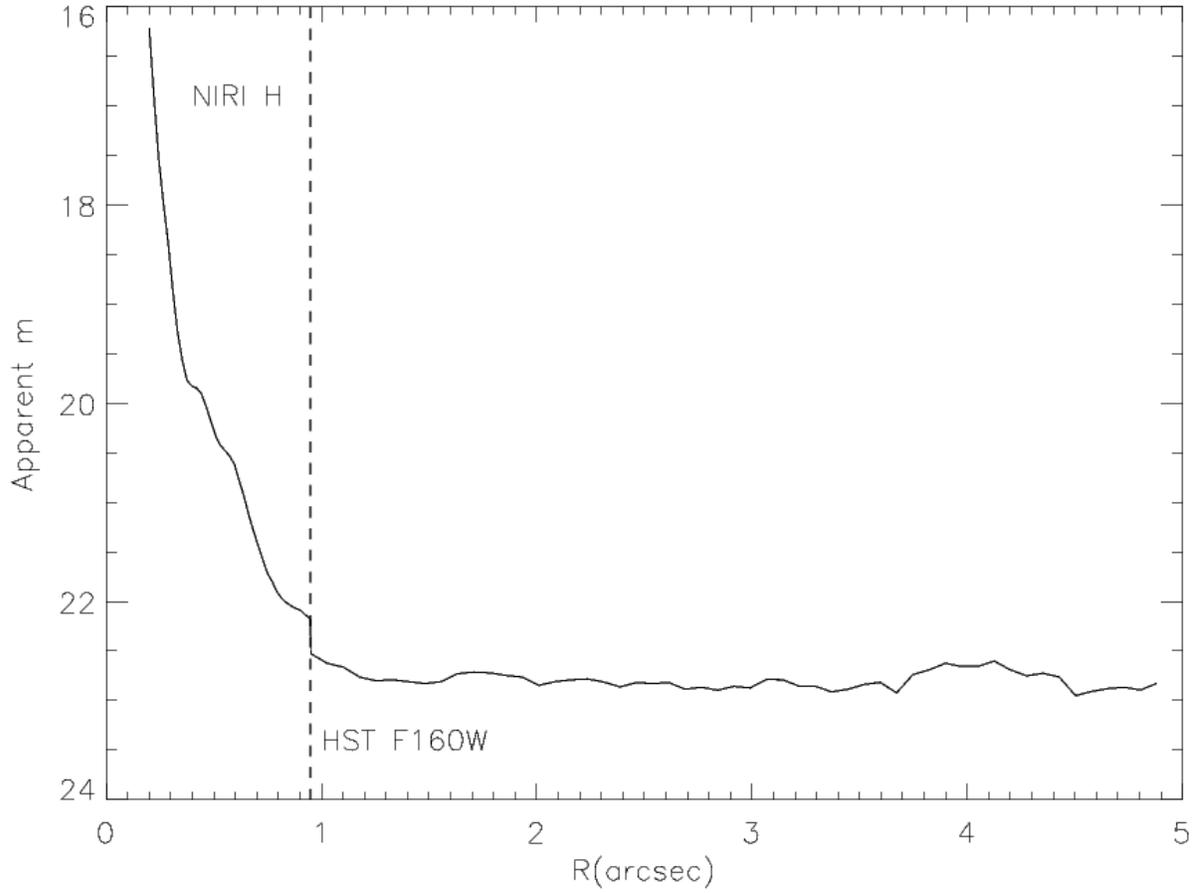}
\caption{\label{fig:sens} The final azimuthally averaged
limiting magnitude curve of our HST and Gemini images.  Our HST observations
were sensitive to objects that had a S/N of $>$ 5 at separations $>$
1\arcsec.  At separations
$<$ 1\arcsec, The Gemini PSF still had significant flux.  To ensure that our sensitivity reflected actual detectability, we used a S/N limit of 10 
$<$ 1\arcsec.}
\end{figure}

Our resulting sensitivity plot in Figure \ref{fig:sens}, incorporating both our Gemini and HST data,
shows the apparent limiting magnitudes in our search from 0.22\arcsec\ to 5\arcsec.  
These results represent the deepest and highest contrast images taken 
around a white dwarf to date. In the NICMOS images beyond 1\arcsec\, our 
sensitivity was limited not by scattered light from G 29-38, but by the 
limited exposure time.

\begin{figure}
\plotone{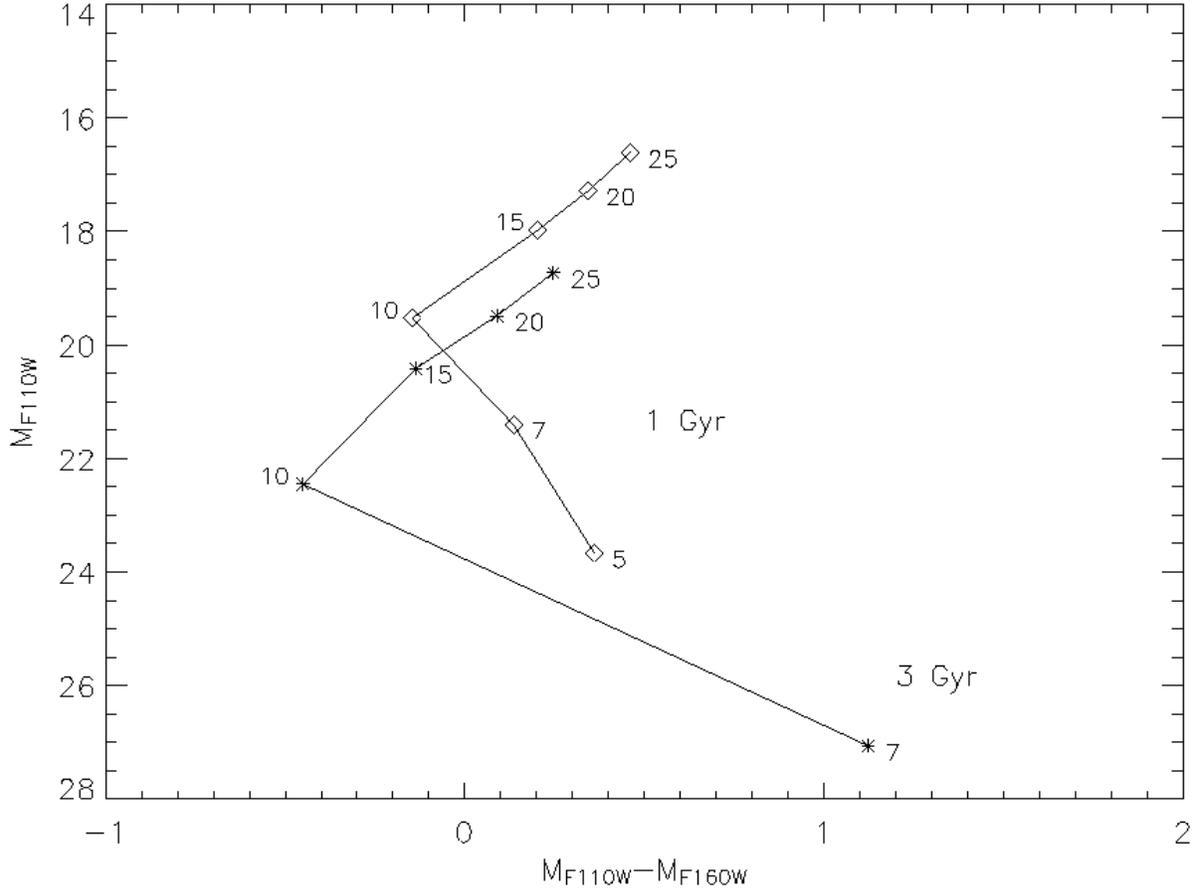}
\caption{\label{fig:cmd} Color-magnitude diagram with isochrones 
of substellar objects with a total age
of between 1 and 3 Gyr in NICMOS filters.  We used the spectral models of
\citet{bsl03} and convolved them with the NICMOS filters.  Numbers on the isochrones refer to the mass in Jupiter masses.  
Numbers 10-25 follow the observed
properties of T dwarfs and have bluer colors.  At effective temperatures of 
$<$ 400 K, water absorption suppresses flux in the F110W filter and again makes
these objects redder.  Colors are sensitive to this absorption and
are uncertain.}
\end{figure}

It is useful to convert the sensitivity in the observed magnitudes or fluxes
into a corresponding companion mass.  Since most
substellar companions do not have long term energy sources, the luminosity
of a brown dwarf or planet that is not significantly insolated is dependent
both on mass and age.  In the present situation we can estimate the age of the
system based on the properties of the host star. For our current sensitivity
 calculation
we chose the most recent models published by \citet{bsl03} and \citet{baraffe03}.  These models 
 are 
difficult to compare to each other and to observations in the near-IR due
 to the presence of H$_2$O molecular 
absorption that can cause variations in predicted magnitudes in different 
photometric systems \citep{stephens04}.  The \citet{baraffe03} magnitudes are
 in the CIT system,
while \citet{bsl03} make their synthetic spectra directly available and thus can be convolved with any filter set.  Both
sets converge to within a magnitude of each other for ages $>$ 1 Gyr
in the J, H, and K filters
but in general, for a given age and mass, the \citet{bsl03} predicted 
magnitudes are fainter.  In Figure \ref{fig:cmd} and our calculations in this
Section,
we use the \citet{bsl03} models.
If the \citet{baraffe03} models are correct, our limits are at most $\sim$1-2
M$_{Jup}$ lower than reported.
In Section \ref{sx.x} we instead use the \citet{baraffe03} models since they
extend to higher mass.

Most models are for ground based J, H, and K filters.  These filters were 
originally designed to avoid atmospheric windows of high near-IR absorption
which is irrelevant for HST filter design.  The wideband NICMOS filters
vaguely resemble their ground-based counterparts, but possess significant 
differences in the case of objects that have deep molecular absorption.
To adequately understand what type of companions one can detect, it is
necessary to take flux calculations from the models and convolve them with
the waveband of interest to get a predicted absolute magnitude for the HST
filters:
\begin{equation}
M_x=-2.5 \log\left(\int \lambda A_\lambda F_\lambda d\lambda\right)+2.5 \log\left(\int \lambda A_\lambda F_{\lambda,Vega} d\lambda\right)
\end{equation}
where $A_\lambda$ is the transmission function of the filter, $F_\lambda$
is the flux of the putative companion, and $F_{\lambda,Vega}$ is the Vega 
flux as calculated by \citet{kurucz}.  This method is preferred for 
detector arrays when calculating synthetic photometry \citep{girardi02}.

Figure \ref{fig:cmd} shows a sample $M_{F110W}$ vs. M$_{F110W}$-M$_{F160W}$ color magnitude plot for
substellar objects  as a function of their mass 
that have ages of 1 Gyr and 3 Gyr \citep{bsl03}.  A 
comparison with \citet{bsl03}'s plots show that the predicted J magnitudes in
their paper and the F110W magnitudes we've calculated
differ by slight amounts due to the different transmission function of the two
filters.  It should also be noted that these predicted 
fluxes are based upon a completely isolated object that is not experiencing
any insolation from its host star.  Companions around WDs would have 
been insolated by their parent star for the main sequence lifetime.   However,
insolation calculations show that this would be insignificant for well
separated companions \citep{burrows04}.  The largest insolation would occur during the 
red giant branch (RGB) and asymptotic giant branch phases (AGB) of post main
sequence evolution.   Calculating the equilibrium temperature
shows that the temperature at 5 AU during these phases would be less than
the temperature experienced by HD 209458B, the Jovian planet in a 0.03 AU 
orbit around a main sequence star.  Insolation of a planet 
during the post main sequence stages of evolution should not be sufficient to 
alter a substellar companion's predicted magnitude from the isolated case.

To get a final prediction of the types of companions to which we are sensitive 
requires a fairly accurate estimate of the WDs total age.  The total age
can be determined from the sum
of a WDs cooling age and its main sequence lifetime.  Estimates of the main 
sequence lifetime can be taken from the initial to final mass ratio 
relationship between WDs and their progenitor stars \citep{weidemann00}.
  Cooling times can be
derived by modeling.  \citet{liebert04}
 gives G29-38's mass and cooling age as 0.7 $\Msun$
and 0.6 Gyr.  Using a theoretical version of the initial-to-final mass
function, M$_i=10.4 \ln \left[(M_{WD}/\Msun)/0.49\right] \Msun $, one derives an initial mass of 3.7 $\Msun$ \citep{wood92}.  The main sequence
 (MS)
lifetime can be estimated by $10 (M/\Msun)^{-2.5}$ Gyr, which gives an MS lifetime of 0.4 Gyr and
thus a total age of 1 Gyr \citep{wood92}.
However, from pulsational studies, the precise mass of G 29-38 is
 0.6 $\Msun$ which, if the 
cooling time remains the same or is a bit longer, leads to an age of 2-3 Gyr
\citep{kleinman98}.  Thus, the age of G 29-38 likely lies between 1 and 3 Gyr.

\subsection{Limits from 2MASS Photometry}
\label{sx.x}

While direct imaging is most sensitive to companions $>$0.2\arcsec\,
unresolved companions could still be present for G29-38.  In
order to rule out companions at separations where imaging or PSF subtraction
could not resolve them, we looked at the near-IR flux of G29-38.   
Low mass companions to WDs have often been discovered
through near-IR excesses \citep{probst82,zuckerman92,green00}.  G29-38 presents a problem due to its 
already well known dust disk, which causes a measurable excess starting at about 1.6\micron.  However, no large excess is predicted for the J band, which we
will use to limit the presence of unresolved substellar companions.  
For our search we use the near-IR photometry of the 2MASS catalogue which has been used in the past to search for 
flux excesses in combination with comparison to model WD atmospheres
\citep{wachter03}.  Using the measured effective 
temperatures, gravities, and distances
 of a WD, we can model the expected
J magnitude (J$_{th}$) using the model atmospheres of \citet{bergeron95}.  
These models cover a wide range of WD effective temperature, gravity, and 
atmospheric composition.  When combined with accurate photometry in the 
visible, these models
can reproduce the flux in the J band 
of a WD to within a few percent \citep{bergeron01}.  The model values of J, H,
and K are based on the CIT filter system, which we converted to 2MASS magnitudes using the color transformations provided by the 2MASS documentation\footnote{
http://www.ipac.caltech.edu/2mass/releases/allsky/doc/sec6\_4b.html}.
Then,
the excess of the expected minus observed J magnitude,
 $\Delta$J=J$_{th}$-J$_{\mbox{2MASS}}$, can be determined.
  An excess of flux in the J band under this notation gives a positive $\Delta$J.  At the
accuracy of 2MASS, limits can be placed on the type of companions present in
close orbit around G29-38.

In order to place robust limits to a J excess for G29-38, we must determine
 the scatter of $\Delta$J from a sample of WDs with
known physical parameters and see what an accurate estimate of a 3$\sigma$
 excess would be.  We would expect the sample to have a median $\Delta$J$\sim$
0 and that the standard deviation of $\Delta$J gives a good estimate of the
1$\sigma$ error in our analysis.  As a demonstration we
 take the sample
of \citet{liebert04}
which includes G29-38 in a study of DA WDs from the PG survey
of UV excess sources.  Of the 374
white dwarfs we chose the brightest 72 of the sample that had
 a J $<$ 15, had unambiguous sources in 2MASS, and had reliable 
photometry, i.e those objects that had quality flags of A or B in the 2MASS
point source catalogue for their J magnitudes.   

If there were a significant number of excesses in the sample then the standard
deviation of the observed minus expected magnitudes
will be overestimated.  Since we cannot {\em a priori} know whether there will
be a large number of excesses or not, we've assumed that there are 
not a significant fraction of WDs with excesses in our samples.
While calculating the standard deviation for each filter, we removed any object
with an excess $>$ 3 $\sigma$ from the sample
 and recalculated the scatter in observed minus
expected magnitudes.  We iterated this process three times.  We found that of 
the 72 sources, only eight objects showed an excess in at least one filter.
These objects are in Table \ref{tab:excesses}.
 
After determining the standard deviation of the sample, we found that
 1$\sigma$\
errors for the sample in the J, H, and K bands were 0.07 mag, 0.1 mag, and
0.15 mag, respectively.  We treated any excesses
greater than 3$\sigma$ as significant, though if an excess was only present in
one band we marked this as a tentative detection.  One exception is G29-38 
itself, which showed only 
a 3.5$\sigma$ excess in the Ks band due to its dust disk, 
which has been amply confirmed in the past.  

Seven objects in our sample showed significant excesses in at least two
filters and one object showed a significant excess only in the Ks band.  These
results are
shown in Table \ref{tab:excesses}.  Of the eight objects, 5 were previously
known (See references in Table \ref{tab:excesses}).   PG~1234+482, PG~1335+369, and PG~1658+441 are new.  
Care was taken to ensure
that the coordinates of new excess candidates 
in the 2MASS fields were correct and that their
 optical photometry was consistent both with that 
reported in \citet{liebert04} and with the distance assumed in the modeling.
The absolute magnitudes of candidate excess 
companions were calculated by taking the excess flux and using the 
distance derived from models of the WDs.  
A spectral type for each excess object was either taken from 
the literature or compared to nearby M and L dwarfs with known distances
\citep{henry94,leggett01}.  The results are presented in Table \ref{tab:excess2}. The spectral types we've determined are rough and need to be 
confirmed through spectroscopic follow-up or high spatial resolution imaging.

PG~1234+482 and PG~1658+441 both were previously studied in the J and K
bands by \citet{green00}
for excesses.  None were reported for either of these objects.
Based on our analysis, PG 1234+482 has significant excesses in the H and Ks filters.  \citet{green00} 
reported a similar K magnitude as that reported in 2MASS but due to larger 
errors in their photometry,
measured it as a marginal excess of $\sim$1.3$\sigma$.  PG~1658+441
 shows only an excess in the Ks 2MASS filter, which is contradicted by
the infrared photometry taken in \citet{green00}.  Their measured magnitude
in K differs by $\sim$0.6 mag from 2MASS, with 
the 2MASS measurements having a higher reported error.
Based on this uncertain photometry, the excess could be due to a
mid L dwarf--the J-K color of such an object would result in a negligible 
excess in J and an observable excess in K$_{s}$ \citep{leggett01}.  This
would be an exciting discovery, if confirmed, as only two substellar objects
are known to orbit nearby white dwarfs \citep{zuckerman88,farihi04}  
PG~1658+441 has been selected and observed
for Program 10255, an HST snapshot program to resolve close
WD+M dwarf binaries.  If an L dwarf is present in an orbit greater than a few
AU, it should be resolved with those observations.

Our resulting 3$\sigma$ limit for G29-38 is then $\Delta$J=0.21, which 
corresponds to an unresolved source with M$_J$=14.8. 
Interpolating from the models of \citep{baraffe03}, the corresponding unresolved companion mass at
 1 and 3 Gyr is 40 $M_{Jup}$ and 58 $M_{Jup}$ 
respectively.

\subsection{Limits from Pulsational Studies}

Claims for the presence of companions around G29-38 have often occurred.
  Its infrared excess was originally attributed to a 
brown dwarf companion, while radial velocity and pulsational timing hinted
at the presence of either a low mass stellar companion or a massive black
hole, all of which were shown to be spurious by more careful, long-term
pulsational timing \citep{kleinman94}.

Pulsational timing is done in a similar fashion to pulsar timing, in that
phase changes of the observed minus calculated (O-C) pulse arrival times
can be used to calculate the projected semi-major axis of the reflex motion
for the white dwarf, $a \sin{i}$.  For
pulsating white dwarfs, the technique requires identifying a
stable pulsational mode and measuring its arrival time very precisely.
Measuring higher derivatives of the period change can also help to further
constrain the Keplerian parameters of a companion orbit before it has
completed a full revolution.  This 
technique for pulsars
has been remarkably effective at finding ``oddball''
planets, such as the first terrestrial 
extrasolar planets ever discovered and a Jovian mass
planet in the metal poor M4 cluster \citep{wolszczan92,sigurdsson03}.

Long baseline timing studies of pulsating white dwarfs can produce very stringent
limits to the types of companions orbiting them, down to tens of Earth masses.
They are limited by the timescale of observations and knowledge of the inclination of the system while probing the inner-most
 orbital separations.  In this sense pulsational timing is generally complementary
to direct imaging searches, the combination of the two providing a comprehensive and sensitive method for searching for extra-solar planets.

\citet{kleinman94} demonstrated that for G29-38, perturbations on the order of
10~s or greater could have been detected around the white dwarf.  In fact, a 
trend was discovered in their data
that had an amplitude of 56~s and a possible period of
8 years.  This was a tentative detection given the possibility of the 
mode that they used being unstable or slowly varying.   However, based on G29-38's parameters, one can estimate
how massive such a companion would be and what
its semi-major axis would be assuming $i\sim$90$^\circ$.  
 Assuming G29-38 has a mass of 0.6 $\Msun$, the derived
minimum
 mass was 21 M$_{Jup}$
with a semi-major axis of 3.4 AU. 
A mass of 0.7 $\Msun$\ does not significantly change these values.  

\begin{figure}
\plotone{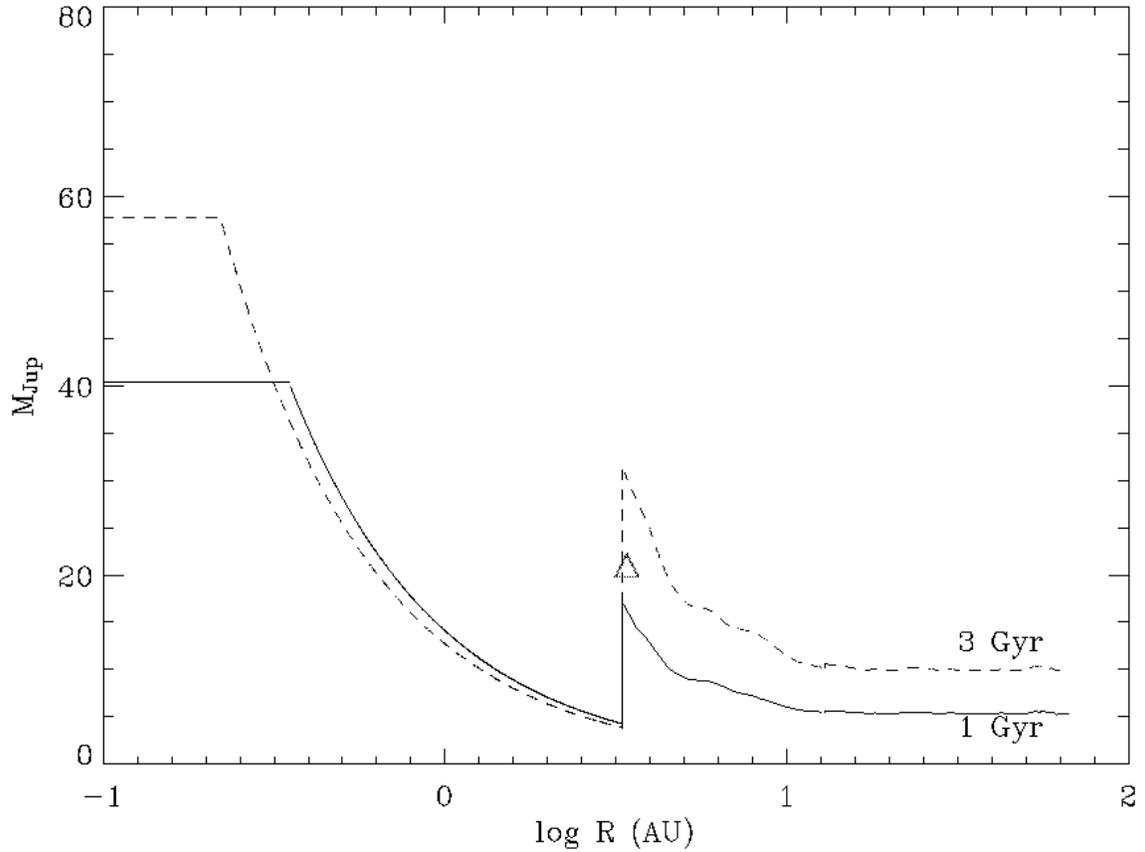}
\caption{\label{fig:finalsens} Combined limits to substellar objects around
G29-38 from a combination of 2MASS photometry, pulsation studies, and our 
high contrast imaging.  The solid and dashed
 lines show the limits for assumed total ages of 1 and 3 Gyr, respectively,
and the triangle shows the expected minimum mass of a companion tentatively 
discovered by pulsational studies.}
\end{figure}

As mentioned above, the noise limit to the \citet{kleinman94} pulsational 
timing allows limits to be placed on the types of companions present with 
orbital timescales of $<$ 8 years.  
Figure \ref{fig:finalsens} shows the combination of the
pulsational timing limits based on the 10~s noise limit and our observational
data.  Our 2MASS photometry limits extend to where the predicted mass
equals that derived from the limits of the pulsational studies, 0.4~AU for
an age of 1 Gyr and 0.2 AU for an age of 3 Gyr. Between those separations and 3~AU, the limits are determined by
the pulsational studies.  Beyond 3~AU the limits are determined by our imaging.
  Overplotted is the separation and mass of the possible companion
detected in the pulsational timing. Our observations weigh against
 the possibility 
of the tentative companion, if the total age of G29-38 is closer to 1 Gyr.  If
the age of G29-38 is closer to 3~Gyr, we can constrain the inclination of the
possible companion's orbit
to be $>$ 44$^\circ$ from face on based on our detection limit of 
30 $M_{Jup}$.  Inspection of the limits shows that any companion $>$ 12 M$_{Jup}$ is ruled out for separations between $\sim$1 AU and 3 AU and $>$ 5 AU if
the age of G 29-38 is close to 1 Gyr.  All but planetary mass objects are ruled out for 
a good portion of the discovery space around this white dwarf.  Further 
observations, such as sensitive radial velocity variations, would provide a 
stronger limit to close in companions than what is possible with 2MASS.

\section{Conclusions}
\label{s4}
We have shown that a combination of high contrast imaging
and photometry of individual relatively young and nearby white dwarfs such as
G 29-38
can effectively probe for high mass planets.
  Information gleaned through this technique we can detect planets
not accessible by other methods.  Any planet discovered could become an 
important spectroscopic target for follow-up.  The information gleaned from a large scale version
of this study may provide key information on planet formation and
evolution in intermediate mass stars as well
as providing a possible explanation for the origin of 
white dwarfs with metal absorption \citep{debes02}.  

If a close companion is involved in the origin of G29-38's dusty disk, 
it must be substellar and if
a well-separated companion is involved it is of planetary mass.  These mass
limits apply if the 
scenario for the formation of DAZs follows \citet{debes02}, where an
 unstable planetary system sends volatile-depleted asteroidal or
cometary material into the inner system.  
The possibility remains that a smaller planet could be 
present.  Indeed, planets of $\sim$1 M$_{Jup}$ or less may be favored
for the DAZ phenomenon \citep[][private communication]{hansen04}.  Planets near
our mass limits may be too efficient at ejecting surviving planetesimals
rather than sending them into the inner system.

Finally, due to the sensitivity of our Gemini observations we can place some 
strong conclusions on previous claims for the presence of close companions due
to pulsational timing by \citet{kleinman94}.  If the age of G29-38 is 1 Gyr, 
we can refute the presence of
a companion at $\sim$3.4 AU.  We can place limits on its mass its
 if the age of G29-38 is closer to 3 Gyr. 
The possibility exists that the companion could be closer to G 29-38 than its 
maximum extent, since the pulsation timing observations were of not sufficient
quality to determine the phase of the initial observations.  
We see no evidence for a companion beyond some 
structure in the AO PSF at a projected separation that does not match 
the predicted orbital separation \citep[][personal communication]{trujillo04}. 

\acknowledgements
We would like to gratefully acknowledge Al Shultz and Glenn Schneider for 
helpful conversations about coronagraphy with NICMOS, and Chad Trujillo and Joe
Jensen for critical help with the inner workings of Altair and the reduction
of Altair imaging data.

Based on observations made with the NASA/ESA Hubble Space Telescope, obtained
 at the Space Telescope Science Institute, which is operated by the 
Association of Universities for Research in Astronomy, Inc., under NASA 
contract NASÊ5-26555. These observations are associated with program \#9834.
Also based on observations obtained at the Gemini Observatory, 
which is operated by the
Association of Universities for Research in Astronomy, Inc., 
under a cooperative agreement
with the NSF on behalf of the Gemini partnership: the National
 Science Foundation (United
States), the Particle Physics and Astronomy Research Council 
(United Kingdom), the
National Research Council (Canada), CONICYT (Chile), the Australian
 Research Council
(Australia), CNPq (Brazil) and CONICET (Argentina). Near-IR Photometry obtained as part of the Two Micron All Sky Survey (2MASS), a joint project of the University 
of Massachusetts and the Infrared Processing and Analysis Center/California 
Institute of Technology, funded by the National Aeronautics and Space 
Administration and the National Science Foundation.  S.S. also acknowledges
funding under the Pennsylvania State University Astrobiology Research Consortium (PSARC).

\bibliography{g29bib}
\bibliographystyle{apj}

\clearpage

\begin{deluxetable}{lccc}
\tablecolumns{4}
\tablewidth{0pc}
\tablecaption{\label{tab:obs} Table of the observations taken of G29-38}
\tablehead{
\colhead{Observation name} & \colhead{Date \& Time(UT)} & \colhead{Filter}
& \colhead{Exposure Time(s)}}
\startdata
N8Q301010 & 2003-10-20 10:07:00 & F205W & 17.942 \\
N8Q301011 & 2003-10-20 10:08:00 & F205W & 17.942 \\
N8Q301020 & 2003-10-20 10:15:20 & F160W & 11.960 \\
N8Q301030 & 2003-10-20 10:20:00 & F110W & 11.960 \\
N8Q304010 & 2003-09-14 19:31:00 & F110W & 575.877 \\
N8Q305010 & 2003-09-14 19:59:00 & F110W & 575.877 \\
N8Q306010 & 2003-09-14 21:07:00 & F160W & 575.877 \\
N8Q307010 & 2003-09-13 21:35:00 & F160W & 575.877 \\
GN-2004A-DD-9 & 2004-08-05 14:81:08 & MKO H & 2220.00 \\
\enddata
\end{deluxetable}

\begin{deluxetable}{ccccccc}
\tablecolumns{7}
\tablewidth{0pc}
\tablecaption{\label{tab:excesses} 2MASS Photometry of PG WDs}
\tablehead{\colhead{PG} & \colhead{J$_{th}$} & \colhead{H$_{th}$} & 
\colhead{K$_{s(th)}$} & \colhead{J} & \colhead{H} & \colhead{K$_s$}
 }
\startdata
0017+061 & 15.33 & 15.49 & 15.56 & 13.74 & 13.19 & 12.98 \\ 
0205+134 & 15.45 & 15.63 & 15.72 & 12.80 & 12.20 & 11.96 \\ 
0824+289 & 14.95 & 15.13 & 15.22 & 12.42 & 11.80 & 11.65 \\
1026+002 & 14.29 & 14.41 & 14.46 & 11.75 & 11.22 & 10.94 \\
1033+464 & 14.93 & 15.08 & 15.17 & 12.56 & 12.03 & 11.75 \\ 
1234+482 & 15.14 & 15.32 & 15.40 & 14.98 & 14.96 & 14.94 \\ 
1335+369 & 15.03 & 15.15 & 15.20 & 13.29 & 12.92 & 12.85 \\ 
1658+441 & 15.26 & 15.40 & 15.50 & 15.44 & 15.53 & 15.05 \\ 
\enddata
\end{deluxetable}

\begin{deluxetable}{cccccc}
\tablecolumns{5}
\tablewidth{0pc}
\tablecaption{\label{tab:excess2} Magnitudes and Spectral Types of 
Excess Candidates}
\tablehead{\colhead{PG} & \colhead{M$_J$} & \colhead{M$_H$} & \colhead{M$_{K_s}$} & 
\colhead{Sp Type} & \colhead{Reference}}
\startdata
0017+061 & 8.98 & 8.29 & 8.05 & M5V & 1\tablenotemark{a} \\
0205+134 & 6.46 & 5.81 & 5.56 & M3.5V & 2 \\
0824+289 & 6.90 & 6.24 & 6.09 & dC+M3V & 3 \\
1026+002 & 8.96 & 8.38 & 8.09 & M5V & 1 \\
1033+464 & 8.15 & 7.56 & 7.26 & M4V & 1 \\
1234+482 & 11.31 & 10.3 & 10.3 & M8V & - \tablenotemark{b} \\
1335+369 & 9.30 & 8.84 & 8.77 & M5.5V & -\tablenotemark{b}  \\
1658+441 &  - & - & 14.1 & L5 & -\tablenotemark{b} \\ 
\enddata
\tablerefs{
(1) \citet{zuckerman92}
(2) \citet{allard94}
(3) \citet{green00}}
\tablenotetext{a}{ \citet{zuckerman92} did not estimate spectral type,
 estimates taken from 2MASS magnitudes of nearby M dwarfs listed in
 \citet{henry94}}
\tablenotetext{b}{This work used 2MASS magnitudes of nearby M dwarfs from 
\citet{henry94} and nearby L, T dwarfs from \citet{leggett01} to determine rough spectral types}

\end{deluxetable}

\clearpage

\clearpage

\clearpage

\clearpage

\end{document}